\documentclass[twocolumn,english,pra,showpacs]{revtex4-1}
\input{Qcircuit.tex}
\usepackage[T1]{fontenc}
\usepackage[latin9]{inputenc}
\usepackage{amsmath}
\usepackage{graphicx}
\usepackage{amssymb}
\usepackage{esint}
\usepackage{epsfig}
\usepackage{amsfonts}
\usepackage{amsmath}
\usepackage{verbatim}
\makeatletter
\@ifundefined{textcolor}{}
{%
 \definecolor{BLACK}{gray}{0}
 \definecolor{WHITE}{gray}{1}
 \definecolor{RED}{rgb}{1,0,0}
 \definecolor{GREEN}{rgb}{0,1,0}
 \definecolor{BLUE}{rgb}{0,0,1}
 \definecolor{CYAN}{cmyk}{1,0,0,0}
 \definecolor{MAGENTA}{cmyk}{0,1,0,0}
 \definecolor{YELLOW}{cmyk}{0,0,1,0}
 }

\makeatletter
\usepackage{dcolumn}
\usepackage{bm}
\usepackage{amsthm}\@ifundefined{definecolor}
 {\@ifundefined{definecolor}
 {\usepackage{color}}{}
}{}

\makeatother

\usepackage{babel}

\makeatother

\usepackage{babel}

\begin{document}
\newtheorem{conjecture}{Conjecture}\newtheorem{corollary}{Corollary}\newtheorem{theorem}{Theorem}
\newtheorem{lemma}{Lemma}
\newtheorem{observation}{Observation}
\newtheorem{definition}{Definition}\newtheorem{remark}{Remark}\global\long\global\long\def\ket#1{|#1 \rangle}
 \global\long\global\long\def\bra#1{\langle#1|}
 \global\long\global\long\def\proj#1{\ket{#1}\bra{#1}}

\title{Non-binary Entanglement-assisted Stabilizer Quantum Codes}

\author{Leng Riguang and Ma Zhi}

 \affiliation{Zhengzhou Information Science and Technology Institute, Zhengzhou 450002, China}

\date{\today}

\begin{abstract}
In this paper, we show how to construct non-binary entanglement-assisted stabilizer quantum codes by using pre-shared entanglement between the sender and receiver. We also give an algorithm to determine the circuit for non-binary entanglement-assisted stabilizer quantum codes and some illustrated examples. The codes we constructed do not require the dual-containing constraint, and many non-binary classical codes, like non-binary LDPC codes, which do not satisfy the condition, can be used to construct non-binary entanglement-assisted stabilizer quantum codes.
\end{abstract}
\maketitle
\section{Introduction}
Errors caused by noises in quantum informational processes are inevitable. One active way of dealing with errors is provided by quantum error-correcting codes~\cite{01,02,03}, which have been found many application in quantum computations and quantum communications, such as the quantum key distributions~\cite{04}, the fault-tolerant quantum computation~\cite{05}and the entanglement purification~\cite{06}. The large majority of work on quantum error-correcting codes has concerned on quantum stabilizers codes~\cite{07,08,09,10,11}and they have become the most widely-used class of quantum error-correcting codes. One reason is that the CSS and CRSS code constructions~\cite{02,07,08,12}allow classical self-orthogonal codes to be easily transformed into quantum stabilizer codes.

Bowen~\cite{13}constructed the first entanglement-assisted quantum error-correcting code from a three-qubit bit-flip code with the help of two pairs of maximally-entangled states. Brun, Devetak and Hsieh~\cite{14,15}showed that if shared entanglement between sender and receiver is available, classical linear quaternary (and binary) codes that are not self-orthogonal can be transformed to entanglement-assisted quantum error-correcting codes. Wilde~\cite{16}gave an algorithm for encoding and decoding a binary entanglement-assisted quantum stabilizer code and Wilede's algorithm not only determine the encoding and decoding circuit for the set of Pauli generators, but also can determine the optimal number of ebits and the measurements the receiver performs to diagnose errors.

In this paper, we consider non-binary entanglement-assisted quantum error-correcting codes on quantum systems which have subsystems of dimension $d=p^m$, where $p$ is a prime. As a shorthand, we can use the term \textquoteleft qudit\textquoteright and quantum codes for qudit systems have been studied, e.g, in~\cite{17,18,19,20,21}. The questions of how to construct, encode and decode qudit entanglement-assisted quantum error-correcting codes have been not explicitly addressed. Here we will present how to use maximally two-particle $d$-dimensional entangled state to construct a qudit entanglement-assisted quantum code and an algorithm to determine how to encode and decode it.

The paper is organized as follows. Section II contains definitions of non-binary quantum states and some qudit quantum gates used later in the paper. Error bases of quantum error-correcting code and qudit stabilizer formalism are introduced in Section III. Section IV first reviews the entanglement-assisted stabilizer formalism, then presents an algorithm to determine the encoding circuit for qudit entanglement-assisted stabilizer code, gives some examples to show how to construct qudit entanglement-assisted stabilizer code in details at last. We discuss our results in Section V.

\section{Non-binary Quantum Systems}
\subsection{Non-binary Quantum States}
Let $d=p^m$ be a power of a prime $p$, $m \geq 1$, $\mathbb{F}_d$ is a finite field with $d$ elements, and let $\mathbb{C}^{d}$ be a $d$-dimensional complex vector space representing the states of a quantum mechanical system. We denote by $|i\rangle$ the vectors of a distinguished orthonormal basis of $\mathbb{C}^{d}$, where the labels $i$ range over the elements of $\mathbb{F}_d$. And the general state of a qudit is given by

$|\phi \rangle= \sum\limits_{i=0}^{d-1} \alpha_{i} |i \rangle$, where $\alpha_{i} \in \mathbb{C}$ and $\sum\limits_{i=0}^{d-1} |\alpha_{i}|^{2} =1$.

Combining several qudits, we obtain a quantum register. The canonical basis states of a quantum register of length $n$ are tensor products of the basis states of the single qudits. For the basis states of a quantum register we use the following notations:
\begin{align*}
&|x_1 \rangle \otimes |x_2 \rangle \otimes \cdots \otimes |x_n \rangle=|x_1 \rangle |x_2 \rangle \cdots |x_n \rangle\\&=|x_{1},x_{2},\cdots ,x_{n} \rangle=|x \rangle.
\end{align*}
A general state of a quantum register of length $n$ is a normalized vector in the exponentially large Hilbert space $\mathbb{H}=(\mathbb{C}^{d})^{\otimes n} \cong \mathbb{C}^{d^n}$, given by

$|\psi \rangle= \sum\limits_{x=0}^{{d^n}-1} \alpha_{x} |x \rangle$, where $\alpha_{i} \in \mathbb{C}$ and $\sum\limits_{x=0}^{{d^n}-1} |\alpha_{x}|^{2} =1$.

\subsection{Qudit Quantum Gates}
In the following we will introduce some qudit gates~\cite{20}on the qudit systems where each qudit correspond to a $q$-dimensional Hilbert space where $q=q^m$ is a prime power.
Let $q$ be a prime power, i. e., $q=p^m$ where $p$ is prime. By $\omega$ we denote a primitive complex $p$-th root of unity, i. e., $\omega=exp(2\pi i/p)$. Furthermore, let $tr(\alpha)$ denote the trace of an element $\alpha \in \mathbb{F}_{q}=\mathbb{F}_{p^m}$ which is defined as $tr(\alpha):= \sum\limits_{i=0}^{m-1}\alpha^{p^i}\in \mathbb{F}_{p}$. When $q=2^m$ and let $B=\{b_{1},\cdots,b_{m}\}$ be an arbitrary self-dual basis of $\mathbb{F}_{q}$ over $\mathbb{F}_{2}$, defining an integer-valued function on $\mathbb{F}_{q}$ as $wgt:\mathbb{F}_{q} \longrightarrow \mathbb{F}_{Z}, \alpha \longrightarrow |\{j:j \in \{1,2,\cdots ,m\}|tr(ab_{j}) \neq 0\}|$. Then define the following operations:
\makeatletter
\renewcommand{\labelenumi}{(\theenumi)}
\renewcommand{\theenumi}{\roman{enumi}}
\makeatother
\begin{enumerate}
\item $X_{\alpha}:=\sum\limits_{x \in \mathbb{F}_{q}} |x+ \alpha \rangle \langle x |$ for $\alpha \in \mathbb{F}_{q}.$

\item $Z_{\beta}:=\sum\limits_{z \in \mathbb{F}_{q}} \omega^{tr(\beta z)} |z \rangle \langle z |$ for $\beta \in \mathbb{F}_{q}.$

\item $DFT:=\frac{1}{\sqrt{q}} \sum\limits_{x,z \in \mathbb{F}_{q}} \omega^{tr(xz)} |x \rangle \langle z |.$

\item $M_{\gamma}:=\sum\limits_{y \in \mathbb{F}_{q}} |\gamma y \rangle \langle y |$ for $\gamma \in \mathbb{F}_{q}\setminus \{0\}.$

\item  $P_{\gamma}:=\sum\limits_{y \in \mathbb{F}_{q}} \omega^{-tr(\frac{1}{2} \gamma y^2)} |y \rangle \langle y |$  for  $q$ is odd;\\
     $P_{\gamma}:=M_{\gamma_{0}}^{-1}\sum\limits_{y \in \mathbb{F}_{q}}{(-i)}^{wgt(y)} |y \rangle \langle y |$, where $\gamma_{0}^{2}=\gamma$ for $q$ is even.

\item $ADD_{ab}:= \sum\limits_{x,y \in \mathbb{F}_{q}} |x \rangle_{a} |x+y \rangle_{b} \langle y |_{b} \langle x |_{a}.$
\end{enumerate}

\section{Qudit Quantum Stabilizer Code}
\subsection{Error Bases}
In order to construct an error-correcting code, one has to specify an error model. The error model can be specified by a set ${\cal E}$ of error operators.
For qudit systems of prime power dimension $q$, we consider the following set of unitary operators: ${\cal E}=\{X_{\alpha}Z_{\beta}:\alpha, \beta\}$. It is not hard to show that those $q^2$ operators are an orthogonal basis with respect to the inner product $\langle A,B \rangle=tr(A^{\dagger}B)$. Furthermore, they generate an error group $G_1$ of size $pq^2$ with center $Z(G_1)=\langle \omega I \rangle$ . Any element of $G_1$ can uniquely be written as $\omega^{\gamma}X_{\alpha}Z_{\beta}$ where $\gamma \in \{0,\dots ,p-1\}$ and $\alpha , \beta \in \mathbb{F}_{q}$. The commutation relations of two elements are $X_{\alpha}Z_{\beta}=\omega^{-tr(\alpha \beta)}Z_{\beta}X_{\alpha}$.

Hence commuting two elements results in a phase factor, i. e.,
\begin{equation}
(X_{\alpha}Z_{\beta})(X_{\alpha^{'}}Z_{\beta^{'}})=\omega^{tr(\alpha^{'} \beta-\alpha \beta^{'})}(X_{\alpha^{'}}Z_{\beta^{'}})(X_{\alpha}Z_{\beta})
\end{equation}
For an $n$-qudit system, the error basis and the error group are the $n$-fold tensor products ${\cal E}^{\otimes n}$ and $G_{n}=:G^{\otimes n}_{1}$, respectively.
\subsection{Qudit Stabilizer Formalism}
The basic idea of stabilizer code is that suppose $S$ is an abelian subgroup of $G_n$, we can define the stabilizer code $C(S)$ associated with $S$ to be $C(S)=\{|\psi \rangle :g|\psi \rangle=|\psi \rangle, \forall g \in S$ . The code $C(S)$ is the subspace fixed by the $S$, and $S$ is called the stabilizer of the code. In other words, the stabilizer code $C(S)$ is defined as the common eigenspace of the operators in $S$.
A group $S$ can be specified by a set of independent generators, $\{g_{i}\}$. These are elements in $S$ that cannot be expressed as products of each other, and such that each element of $S$ can be written as a product of elements from the set. The benefit of using generators is that it provides compact representation of the group and to see whether a particular vector $| \phi \rangle$ is stabilized by a group $S$, we need only to check whether $| \phi \rangle$ is stabilized by these generators of $S$.
Any element $E$ of the error group $G_n$ can uniquely be written as $E=\omega^{\gamma}(X_{\alpha_{1}}Z_{\beta_{1}}) \otimes (X_{\alpha_{2}}Z_{\beta_{2}}) \otimes \cdots \otimes (X_{\alpha_{n}}Z_{\beta_{n}})=:\omega^{\gamma}X_{\alpha}Z_{\beta}$, where $\omega \in \{0, \cdots ,p-1\}$ and $\alpha=(\alpha_{1},\alpha_{2}, \cdots \alpha_{n}), \beta=(\beta_{1},\beta_{2}, \cdots \beta_{n}) \in \mathbb{F}^{n}_{q}$. The weight of an element $X_{\alpha}Z_{\beta}$ is the number of indices $i$ for which not both $\alpha_{i}$ and $\beta_{i}$ are zero. From the commutation relation (1), it follows that for $(\alpha, \beta)(\alpha^{'}, \beta^{'}) \in \mathbb{F}^{n}_{q} \times \mathbb{F}^{n}_{q}$,
\begin{equation}
(X_{\alpha}Z_{\beta})(X_{\alpha^{'}}Z_{\beta^{'}})=\omega^{(\alpha, \beta) \ast (\alpha^{'},\beta^{'})}(X_{\alpha^{'}}Z_{\beta^{'}})(X_{\alpha}Z_{\beta})
\end{equation}
where the inner product $\ast$ is defined by
\begin{equation}
(\alpha, \beta) \ast (\alpha^{'},\beta^{'})=\sum\limits_{i=1}^{n}tr(\alpha^{'}_{i} \beta_{i}-\alpha \beta^{'}_{i}).
\end{equation}
This shows that the group $\overline{G_{n}}:=G_{n} / \langle \omega I \rangle$ is isomorphic to $\mathbb{F}^{n}_{q} \times \mathbb{F}^{n}_{q}$. we define the symplectic product of two elements $g=X_{\alpha}Z_{\beta}$ and $g^{'}=X_{\alpha^{'}}Z_{\beta^{'}}$ is $g \odot g^{'}=(\alpha, \beta) \ast (\alpha^{'},\beta^{'})$. And two elements $g$ and $g_{'}$ commute if and only if $g \odot g^{'}=0$. Let $\{g_{1},g_{2},\cdots, g_{n-k} \}$ where $g_{i}=\omega^{\gamma_{i}}X_{\alpha_{i}}Z_{\beta_{i}}$ with $\gamma_{i} \in \{0,\cdots,p-1 \}$ and $(\alpha_{i},\beta_{i}) \in \mathbb{F}^{n}_{q} \times \mathbb{F}^{n}_{q}$ be a minimal set of generators for $S$ which is an abelian subgroup of $G_{n}$. Then we can write a stabilizer matrix of the corresponding stabilizer code $C(S)$ in the form
$$
\left(
\begin{array}{c|c}
\alpha_1 & \beta_1\\
\alpha_2 & \beta_2\\
\vdots & \vdots\\
\alpha_{n-k} & \beta_{n-k}
\end{array}
\right)\in \mathbb{F}_q^{(n-k)\times 2n}.
$$
Any error operator $E$ that does not commute with all elements $g \in S$ will change the eigenvalue of an eigenstate $| \phi \rangle$ of $S$ which can be detected by a measurement. But if $E \in Z(S)-S$, where $Z(S)$ is the centralizer of $S$, then $E$ changes elements of $C(S)$ but does not take them out of $C(S)$. So $E$ will be an undetectable error for this code. And a stabilizer code $C(S)$ can correct a set of errors ${\cal E}$ if and only if $E^{\dagger}_{1}E_{2} \in S \cup (G_{n}-Z(S))$ for all $E_{1},E_{2} \in {\cal E}$.
\subsection{Clifford Encoding Unitary}
We will use the qudit quantum gates which are Clifford operations to encode and decode qudit stabilizer codes, and briefly comment on these encoding operations.

The matrix for Fourier gate $DFT$ acting on a single qudit is $\overline{DFT}:=\left[\begin{smallmatrix}0&-1\\1&0\end{smallmatrix}\right]$.

The matrix for Multiplier gate $M_{\gamma}$ acting on a single qudit is $\overline{M}_\gamma:=\left[\begin{smallmatrix}\gamma^{-1}&0\\0&\gamma\end{smallmatrix}\right]$.

The matrix for Phase gate $P_{\gamma}$, neither $q$ is odd nor $q$ is even, acting on a single qudit is $\overline{P}_\gamma:=\left[\begin{smallmatrix}1&\gamma\\0&1\end{smallmatrix}\right]$.

The matrix for $ADD$ gate acting on two single qudits is $\overline{ADD}_{ab}:=\left[\begin{smallmatrix}1&0&-1&0\\0&1&0&0\\ 0&0&1&0\\0&1&0&1 \end{smallmatrix}\right]$.

The Fourier gate DFT transforms the error basis under conjugation as follows: $Z_{\beta} \longrightarrow X_{\beta}, X_{\alpha} \longrightarrow Z_{-\alpha}.$

The Multiplier gate $M_{\gamma}$ transforms the error basis under conjugation as follows: $X_{\alpha} \longrightarrow X_{\gamma^{-1} \alpha}, Z_{\beta} \longrightarrow Z_{\gamma \beta}.$

The Phase gate $P_{\gamma}$ transforms the error basis under conjugation as follows: $X_{\alpha} \longrightarrow X_{\alpha}Z_{\alpha \gamma}, Z_{\beta} \longrightarrow Z_{\beta}.$

For the ADD gate, the first qudit is the "control" qudit and the second qudit is the "target" qudit. The ADD gate transforms the error basis under conjugation as follows: $X_{\alpha} \otimes Z_{\beta} \longrightarrow X_{\alpha}Z_{\beta} \otimes X_{-\alpha}Z_{\beta}.$

The next chapter will detail an algorithm that determines a Clifford encoding circuit for qudit entanglement-assisted stabilizer code.

\section{Qudit Entanglement-assisted Stabilizer Quantum Code}
\subsection{The Entanglement-assisted Stabilizer Formalism}
The entanglement-assisted stabilizer formalism is a significant extension of the standard stabilizer formalism that incorporates shared entanglement as a resource for protecting quantum information[14,15]. The advantage of entanglement-assisted stabilizer codes is that the sender's operators do not necessarily have to form an abelian subgroup. The sender can make clever use of her shared entangled pairs so that the global stabilizer is abelian and thus forms a valid quantum error-correcting code.

The entangled state we use is a maximal two-particle $d$-dimensional entangled state and We express the entangled state $|\Phi^{+} \rangle$ shared between a sender Alice and a receiver Bob as follows: $|\Phi^{+} \rangle \equiv \sum\limits_{k=0}^{d-1}|k \rangle |k \rangle / \sqrt{d}$.
The two operators $X_{1}^{A}X_{1}^{B}$ and $Z_{1}^{A}Z_{p-1}^{B}$ can stabilize this entangled state. These two operators commute: $X_{1}^{A}X_{1}^{B} \odot Z_{1}^{A}Z_{p-1}^{B}=0$, but the local operators do not commute: $X_{1}^{A} \odot Z_{1}^{A}=1$ , $X_{1}^{B} \odot Z_{p-1}^{B}=p-1$. The above communication relations hint at a way that we can resolve noncommutativity in a set of generators.

Now we introduce the general construction of an entanglement-assisted code. Suppose that there is a nonabelian subgroup $S$ of size $2c+a$, if there exists a minimal set of independent generators $\{\overline{Z}^{1}, \cdots \overline{Z}^{c+a},\overline{X}^{a+1} \cdots \overline{X}^{a+c}\}$ for $S$ with the following commutation ralations:

 \begin{equation}
\begin{aligned}
 &\forall i,j \quad \overline{Z}^{i} \odot \overline{Z}^{j}=0, \quad \forall i,j \quad \overline{X}^{i} \odot \overline{X}^{j}=0 \\
 &\forall i \neq j \quad \overline{X}^{i} \odot \overline{Z}^{j}=0, \quad \forall i \quad \overline{X}^{i} \odot \overline{X}^{i}=0
\end{aligned}
\end{equation}

then there exists an $[[n,k;c]]$ entanglement-assisted code that employs $c$ entangled qudits and $a$ ancilla qudits to encode $k$ information qudits. And the decomposition of $S$ into the above minimal generating set determines that the code requires $a$ ancilla qudits and $c$ entangled qudits, the parametes $a$ and $c$ generally depend on the set of generators in $S$ and the number of encoded qudits $k$ is equal to $n-a-c$.

And we can also partition the nonabelian group $S$ into two subgroups: the isotropic subgroup $S_{I}$ and the entanglement subgroup $S_{E}$. The isotropic subgroup $S_{I}$ is a commuting subgroup of $S$ and thus corresponds to ancilla qudits: $S_{I}:=\{ \overline{Z}^{1}, \cdots \overline{Z}^{a}\}$. The elements of the entanglement subgroup $S_{E}$ come in noncommuting pairs and thus correspond to halves of entangled qudits: $S_{E}:=\{ \overline{X}^{a+1}, \cdots \overline{X}^{a+c},\overline{Z}^{a+1}, \cdots \overline{Z}^{a+c}\}$. The two subgroups $S_{I}$ and $S_{E}$ play a role in the error-correcting conditions for the entanglement-assisted stabibizer formalism. An entanglement-assisted code corrects errors in a set ${\cal E}$ if $\forall E_{1},E_{2} \in {\cal E} \quad E^{\dagger}_{1}E_{2} \in S_{I}$ or $E^{\dagger}_{1}E_{2} \in G_{n}- Z(\langle S_{I},S_{E} \rangle)$ .

The conditions correspond to error pairs $E_{1},E_{2}$ in an error set ${\cal E}$. The first condition corresponds to the passive error-correcting capability of the code, and the second condition corresponds to its active error-correcting capability.

 The operation of an $[[n,k;c]]$ entanglement-assisted stabilizer quantum code have the following steps.

(i). The sender and receiver share $c$ entangled qudits before quantum communication begins and the sender has $a$ ancilla qudits. The unencoded state is a simultaneous $+1$-eigenstate of the following operators:
\begin{equation}
\begin{aligned}
 &\{ {X}_{1}^{a+1} | {X}_{1}^{1}, \cdots, {X}_{1}^{a+c} | {X}_{1}^{c},{Z}_{1}^{a+1} | {Z}_{p-1}^{1},\cdots,  \\&{Z}_{1}^{a+c} | {Z}_{1}^{c}, Z_{1}^{1},\cdots Z_{1}^{a}\}.
\end{aligned}
\end{equation}
The operators to the right of the vertical bars indicate the receiver's half of the shared entangled qudits. The sender encodes her $k$ information qudits with the help of $a$ ancilla qudits and her half of the $c$ entangled qudits. The encoding unitary transforms the unencoded operators to the following encoded operators:
\begin{equation}
\begin{aligned}
&\{ \overline{X}_{1}^{a+1} | {X}_{1}^{1}, \cdots, \overline{X}_{1}^{a+c} | {X}_{1}^{c}, \overline{Z}_{1}^{a+1} | {Z}_{p-1}^{1},\cdots, \\& \overline{Z}_{1}^{a+c} | {Z}_{1}^{c}, \overline{Z}_{1}^{1},\cdots \overline{Z}_{1}^{a}\}.
\end{aligned}
\end{equation}
(ii). The sender sends her $n$ qudits over a noisy quantum communication channel. The noisy channel affects these $n$ qudits only and does not affect the receiver's half of the $c$ entangled qudits.

(iii). The receiver combines his half of the $c$ entangled qudits with those he receives from the noisy quantum channel. He performs measurements on all $n+c$ qudits to diagnose an error that may occur on the $n$ qudits.

(iv). After estimating which error occurs, the receiver performs a recovery operation that reverses the estimated error.
\subsection{Algorithm}
 In this section we derive an encoding algorithm for entanglement-assisted stabilizer quantum code over qudit systems of prime power dimension $d=p^m$. The main idea is to see whether a nonabelian stabilizer group $S \subseteq G_{n}$ of the entanglement-assisted stabilizer quantum code $C(S)=[[n,k;c]]_q$ is isomorphic to $S_{0}:=\{X_{1}^{1},\cdots, X_{1}^{c}, Z_{1}^{1},\cdots, Z_{1}^{c},Z_{1}^{c+1},\cdots, Z_{1}^{n-k-c} \}$ for which encoding is particularly easy. If it is isomorphic, then the nonabelian stabilizer group $S$ and $S_{0}$ are conjugated to each other there exists a transformation $D$ such that $D_{-1}SD=S_{0}$.

 The algorithm consists of row and column operations on the Check matrix. Row operations do not affect the error-correcting properties of the code but are crucial for arriving at the optimal decomposition from the fundamental theorem of symplectic geometry. The operations available for manipulating columns of the check matrix are the above operations. The operations have the following effects on entries in the binary matrix:
\begin{enumerate}
\item A Fourier gate on qudit $i$ swaps multiply column $i$ by $-1$ in the $X$ matrix with column $i$ in the $Z$ matrix.

\item A Multiplier gate on qudit $i$ multiplies column $i$ by invertible integer $q_{-1}$ in the $X$ matrix and multiplies column $i$ by invertible integer $q$ in the $Z$ matrix.

\item A Phase gate on qudit $i$ adds $\gamma$ times column $i$ in the $X$ matrix to column $i$ in the $Z$ matrix.

\item A $ADD$ gate from qudit $i$ to qudit $j$ subtracts column $i$ from column $j$ in the $X$ matrix and adds column $j$ to column $i$ in the $Z$ matrix.
\end{enumerate}

Before introducing our algorithm, we first give a theorem which the first step will use.

Theorem 1. Let $g_1=X_{\alpha_{1}}Z_{\beta_{1}}$,$g_2=X_{\alpha_{2}}Z_{\beta_{2}}$,$\cdots$,$g_{n-k}=X_{\alpha_{n-k}}Z_{\beta_{n-k}}$ are independent generators in the check matrix and $d=p^m$ is a prime power. If there at least exists one pair of generators do not commute, for convenience, let $g_{1}$ and $g_{3}$ do not commute and $g_{1} \odot g_{2}=a_{1},g_{1} \odot g_{3}=a_{2},a_{1} \in \mathbb{F}_{p}, a_{2} \in \mathbb{F}_{p}^{\ast}$, then there must exist an integer $m$, such that $g_{1} \odot (g_{2}g_{3}^{m})=1 \Longleftrightarrow (\alpha_{1},\beta_{1})\ast [ (\alpha_{2},\beta_{2})+m(\alpha_{3},\beta_{3})]=1$.

Proof: we only need to prove that there must exists an $i^{\ast}$ such that $a_{2}|p-a_{1}+1+i^{\ast}p$. Seeking a contradiction, if $a_{2} \nmid p-a_{1}+1+ip \Longleftrightarrow p-a_{1}+1+ip$ mod $a_2 \neq 0$ for all $0 \leq i \leq q-1$, there at least exists $i_{1}$ and $i_{2}$ such that $p-a_{1}+1+i_{1}p \equiv p-a_{1}+1+i_{2}p$ mode $a_{2} \Longleftrightarrow (i_{1}-i_{2})p \equiv 0$ mod $a_{2}$ because there are $a_{2}$ integers. But $0 \leq i_{1},i_{2}< a_{2}$, $p$ is a prime, so it is a contradiction. We let $m=(p-a_{1}+1+i_{\ast}p)/2$, then $(\alpha_{1},\beta_{1})\ast [(\alpha_{2},\beta_{2})+m(\alpha_{3},\beta_{3})]=a_{1}+p-a_{1}+1+i^{\ast}p=1$. And it completes the proof.

Now we introduce our algorithm for determining an encoding circuit and the optimal number of ebits for the qudit entanglement-assisted code. And the algorithm can be divided into two main steps.

(i).$m \leftarrow 0$, $i$ from $2m$ to $n-k$, compute the symplectic inner products between $i$ row and $i+1, \dots , n-k$ rows. If all the symplectic inner products are zero, leave the matrix as it is and go to (ii). Otherwise, let the symplectic inner product $i$ row and $j$ row is not zero, use the method of theorem 1 to make that the product between $i$ row and $i+1$ row is one. Then arrange the $i,\cdots, n-k$ rows in the top of the matrix and the $1, \cdots i-1$ rows in the bottom and $m \leftarrow m+1$. Use Fourier operation, Multiplier operation, Phase operation, ADD operation or combinations of these operations to achieve the $m$ entry in the $X$ matrix of the $2m-1$ row is one, other entries of the $2m-1$ row are zero and the $m$ entry in the $Z$ matrix of the $2m$ row is one, other entries of the $2m$ row are zero. Then add $2m-1$ and $2m$ rows to $2m+1, \cdots , n-k$ rows so that the $m$ entries in the $X$ matrix and in the $Z$ matrix of the $2m+1, \cdots ,n-k$ rows are zero.

(ii).If $2m<n-k-2$, $i$ from $2m+1$ to $n-k$, Use Fourier operation, Multiplier operation, Phase operation, ADD operation or combinations of these operations and row operations to achieve the $i-m$ entry in the $Z$ matrix of the $i$ row is one, other entries of the $i$ row are zero. If $2m=n-k-2$ compute the symplectic inner product between $n-k-1$ row and $n-k$ row, if the symplectic inner product is one, use Fourier operation, Multiplier operation, Phase operation, ADD operation or combinations of these operations to achieve the $(n-k)/2$ entry in the $X$ matrix of the $n-k-1$ row is one , other entries of the $n-k-1$ row are zero and the $(n-k)/2$ entry in the $Z$ matrix of the $n-k$ row is one, other entries of the $n-k$ row are zero; if the symplectic inner product is zero, Use the relevant operations to achieve the $m+1$ entry in the $Z$ matrix of the $n-k-1$ row is one, other entries of the row $n-k-1$ are zero and the $m+2$ entry in the $Z$ matrix of the $n-k$ row is one, other entries of the row $n-k$ are zero; if the symplectic inner product is not one or zero, fail. If $2m=n-k-2$, Use the relevant operations to achieve the $m+1$ entry in the $Z$ matrix of the $n-k$ row is one, other entries of the row $n-k$ are zero.

\subsection{Examples}
(i).The first example is a group $S$ generated by a non-commuting set of operators over $\mathbb{F}_5$ and its checking matrix is
$$
(X|Z)=\left[\begin{array}{c|c}
3 \quad 1 \quad 1 \quad 0 & 1 \quad 2 \quad 0 \quad 2\\
0 \quad 3 \quad 0 \quad 4 & 2 \quad 4 \quad 1 \quad 3\\
1 \quad 1 \quad 0 \quad 2 & 3 \quad 1 \quad 1 \quad 2\\
2 \quad 3 \quad 1 \quad 0 & 4 \quad 0 \quad 1 \quad 3
\end{array}\right]
$$
The algorithm begins by computing the symplectic inner product between the first row and all other rows. And the symplectic inner product between the first row and the second row is 2, the symplectic inner product between the first row and the three row is 4, use the method of lemma 3, we get $m=1$, we add the three row to the second row so that the symplectic inner product between the first row and the second row is 1. The matrix becomes
$$
(X|Z)=\left[\begin{array}{c|c}
3 \quad 1 \quad 1 \quad 0 & 1 \quad 2 \quad 0 \quad 2\\
1 \quad 4 \quad 0 \quad 1 & 0 \quad 0 \quad 2 \quad 0\\
1 \quad 1 \quad 0 \quad 2 & 3 \quad 1 \quad 1 \quad 2\\
2 \quad 3 \quad 1 \quad 0 & 4 \quad 0 \quad 1 \quad 3
\end{array}\right]
$$
Perform Multiply, ADD or combinations of both operations to achieve the leftmost entry in the first row of the $X$ matrix, perform ADD operations to clear the entries in the $X$ matrix. Proceed to the clear the entries in the first row of the $Z$ matrix. Perform phase operations to clear the leftmost entry in the row of the $Z$ matrix if it is not equal to zero. Then use DFT and ADD operations to clear the other entries in the first row of the $Z$ matrix.
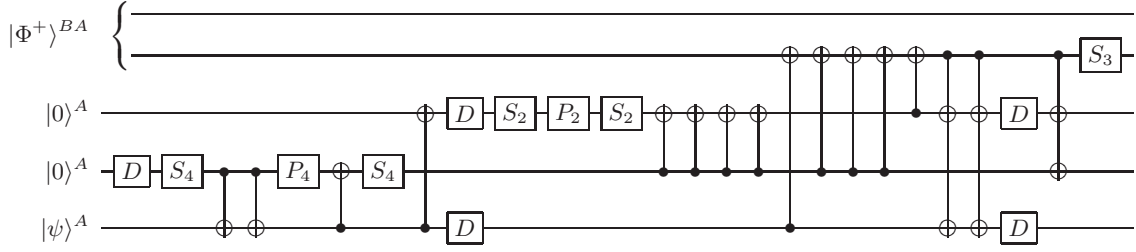
\begin{figure*}
[ptb]
\begin{center}
\[
\Qcircuit@C=0.5em @R=1.0em  {
&  &  & \qw &\qw  &\qw  &\qw  &\qw  & \qw &\qw &\qw  &\qw &\qw &\qw &\qw &\qw  &\qw  &\qw  & \qw& \qw& \qw& \qw& \qw& \qw& \qw& \qw& \qw& \qw& \qw\gategroup{1}%
{3}{2}{3}{1.0em}{\{}\\
& \lstick{\raisebox{2em}{$\ket{\Phi^{+}}^{BA}$}}&  &\qw  &\qw  & \qw &\qw  &\qw  &\qw  &\qw  &\qw  & \qw  &\qw  &\qw & \qw  & \qw & \qw &\qw & \targ \qwx[3]& \targ \qwx[2] & \targ \qwx[2]& \targ \qwx[2]
& \targ \qwx[1] & \ctrl{1} \qwx[3] & \ctrl{1} \qwx[3] & \qw & \ctrl{1} \qwx[2] & \gate{S_{3}} & \qw\\
& \lstick{\ket{0}^A}    & \qw &\qw & \qw & \qw & \qw &\qw &\qw & \targ \qwx[2] & \gate{D} & \gate{S_{2}} & \gate{P_{2}} & \gate{S_{2}} & \targ \qwx[1] & \targ \qwx[1] & \targ \qwx[1]  & \targ \qwx[1] & \qw& \qw & \qw & \qw & \control \qw & \targ & \targ& \gate{D} & \targ  & \qw & \qw\\
& \lstick{\ket{0}^A}    & \gate{D} & \gate{S_{4}} &\ctrl{1} \qwx[1] &\ctrl{1} \qwx[1]&\gate{P_{4}} &\targ \qwx[1] &\gate{S_{4}} & \qw & \qw & \qw & \qw & \qw &\control \qw &\control \qw &\control \qw  & \control \qw & \qw & \control \qw& \control \qw & \control \qw
& \qw& \qw& \qw& \qw & \targ  & \qw& \qw\\
& \lstick{\ket{\psi}^A}   &\qw &\qw &\targ &\targ &\qw & \control \qw &\qw & \control \qw & \gate{D} & \qw & \qw & \qw & \qw & \qw & \qw & \qw& \control \qw & \qw & \qw& \qw& \qw
& \targ& \targ & \gate{D} & \qw& \qw& \qw}
\]
\end{center}
\caption{Encoding circuit for the entanglement-assisted code with parameters $[[4,1;1]]_5$. The ``D'' gate is a Fourier gate, the ``M'' gate is a Multiplier gate and the ``P'' gate is a Phase gate.}
\label{1}
\end{figure*}

For our example, perform $S_{3}$ on qudit one, then perform ADD from qudit one to qudit two and from qudit one to qudit three. The matrix becomes
$$
(X|Z)=\left[\begin{array}{c|c}
1 \quad 0 \quad 0 \quad 0 & 0 \quad 2 \quad 0 \quad 2\\
2 \quad 2 \quad 3 \quad 1 & 2 \quad 0 \quad 2 \quad 0\\
2 \quad 4 \quad 3 \quad 2 & 0 \quad 1 \quad 1 \quad 2\\
4 \quad 4 \quad 2 \quad 0 & 0 \quad 0 \quad 1 \quad 3
\end{array}\right]
$$
Perform DFT on qudit two and qudit four, then perform ADD from qudit one to qudit two with two times and from qudit one to qudit four with two times. The matrix becomes
$$
(X|Z)=\left[\begin{array}{c|c}
1 \quad 0 \quad 0 \quad 0 & 0 \quad 0 \quad 0 \quad 0\\
2 \quad 1 \quad 3 \quad 1 & 1 \quad 3 \quad 2 \quad 4\\
2 \quad 2 \quad 3 \quad 3 & 3 \quad 1 \quad 1 \quad 3\\
4 \quad 2 \quad 2 \quad 0 & 2 \quad 1 \quad 1 \quad 0
\end{array}\right]
$$
The first row is complete. We now proceed to clear the entries in the second row. Perform ADD from qudit two to qudit one with two times, from qudit three to qudit one with three times and from qudit four to qudit one. The matrix becomes
$$
(X|Z)=\left[\begin{array}{c|c}
1 \quad 0 \quad 0 \quad 0 & 0 \quad 0 \quad 0 \quad 0\\
0 \quad 1 \quad 3 \quad 1 & 1 \quad 0 \quad 0 \quad 0\\
1 \quad 2 \quad 3 \quad 3 & 3 \quad 2 \quad 0 \quad 1\\
4 \quad 2 \quad 2 \quad 0 & 2 \quad 0 \quad 2 \quad 2
\end{array}\right]
$$
Perform DFT on qudit two, qudit three and qudit four, then perform ADD from qudit two to qudit one, from qudit three to qudit one with three times and from qudit four to qudit one. The matrix becomes
$$
(X|Z)=\left[\begin{array}{c|c}
1 \quad 0 \quad 0 \quad 0 & 0 \quad 0 \quad 0 \quad 0\\
0 \quad 0 \quad 0 \quad 0 & 1 \quad 0 \quad 0 \quad 0\\
3 \quad 2 \quad 0 \quad 1 & 3 \quad 1 \quad 1 \quad 0\\
1 \quad 0 \quad 2 \quad 2 & 2 \quad 0 \quad 4 \quad 2
\end{array}\right]
$$
The first two rows are now complete. They need one ebit to compensate for their noncommutativity or nonorthogonality with respect to the symplectic inner product.

Now we perform row operations that are similar to the "symplectic Gram-Schmidt orthogonalization" .Add a multiple of row one to any other row that is not zero as the leftmost entry in its $X$ matrix so that its leftmost entry in its $X$ matrix is zero. Add a multiple of row two to any other row that is not zero as the leftmost entry in its $Z$ matrix so that its leftmost entry in its $Z$ matrix is zero. For our example, we add a two multiple of row one to row three, a four multiple of row one to row four and a two multiple of row two to row three, a three multiple of row two to row four. The matrix becomes
$$
(X|Z)=\left[\begin{array}{c|c}
1 \quad 0 \quad 0 \quad 0 & 0 \quad 0 \quad 0 \quad 0\\
0 \quad 0 \quad 0 \quad 0 & 1 \quad 0 \quad 0 \quad 0\\
0 \quad 2 \quad 0 \quad 1 & 0 \quad 1 \quad 1 \quad 0\\
0 \quad 0 \quad 2 \quad 2 & 0 \quad 0 \quad 4 \quad 2
\end{array}\right]
$$
The first two rows are now symplectically orthogonal to all other rows.

We know that the last two rows are symplectically orthogonal to each other from computing their symplectic inner product. Perform ADD from qudit three to qudit two with four times, perform $S_{2}$ on qudit two, then $P_{2}$ on qudit two. The matrix becomes
$$
(X|Z)=\left[\begin{array}{c|c}
1 \quad 0 \quad 0 \quad 0 & 0 \quad 0 \quad 0 \quad 0\\
0 \quad 0 \quad 0 \quad 0 & 1 \quad 0 \quad 0 \quad 0\\
0 \quad 1 \quad 0 \quad 1 & 0 \quad 0 \quad 0 \quad 0\\
0 \quad 1 \quad 2 \quad 2 & 0 \quad 3 \quad 4 \quad 2
\end{array}\right]
$$
Perform $S_{2}$ on qudit two, Perform DFT on qudit two and qudit four, then perform ADD from qudit four to qudit two. The matrix becomes
$$
(X|Z)=\left[\begin{array}{c|c}
1 \quad 0 \quad 0 \quad 0 & 0 \quad 0 \quad 0 \quad 0\\
0 \quad 0 \quad 0 \quad 0 & 1 \quad 0 \quad 0 \quad 0\\
0 \quad 0 \quad 0 \quad 0 & 0 \quad 1 \quad 0 \quad 0\\
0 \quad 0 \quad 2 \quad 2 & 0 \quad 1 \quad 4 \quad 4
\end{array}\right]
$$
Add a four multiple of row three to row four, then perform $S_{4}$ on qudit three, perform ADD from qudit four to qudit three, $P_{4}$ on qudit three. The matrix becomes
$$
(X|Z)=\left[\begin{array}{c|c}
1 \quad 0 \quad 0 \quad 0 & 0 \quad 0 \quad 0 \quad 0\\
0 \quad 0 \quad 0 \quad 0 & 1 \quad 0 \quad 0 \quad 0\\
0 \quad 0 \quad 0 \quad 0 & 0 \quad 1 \quad 0 \quad 0\\
0 \quad 0 \quad 1 \quad 2 & 0 \quad 0 \quad 0 \quad 0
\end{array}\right]
$$
Perform ADD from qudit three to qudit four with two times, then perform $S_{4}$ on qudit three and DFT on qudit three. The matrix becomes
$$
(X|Z)=\left[\begin{array}{c|c}
1 \quad 0 \quad 0 \quad 0 & 0 \quad 0 \quad 0 \quad 0\\
0 \quad 0 \quad 0 \quad 0 & 1 \quad 0 \quad 0 \quad 0\\
0 \quad 0 \quad 0 \quad 0 & 0 \quad 1 \quad 0 \quad 0\\
0 \quad 0 \quad 0 \quad 0 & 0 \quad 0 \quad 1 \quad 0
\end{array}\right]
$$
Adding one entangled qudit to resolve the anticommutativity of the first two generators, the matrix becomes
$$
(X|Z)=\left[\begin{array}{c|c}
1 \quad 0 \quad 0 \quad 0 \quad 1 & 0 \quad 0 \quad 0 \quad 0 \quad 0\\
0 \quad 0 \quad 0 \quad 0 \quad 0 & 1 \quad 0 \quad 0 \quad 0 \quad 4\\
0 \quad 0 \quad 0 \quad 0 \quad 0 & 0 \quad 1 \quad 0 \quad 0 \quad 0\\
0 \quad 0 \quad 0 \quad 0 \quad 0 & 0 \quad 0 \quad 1 \quad 0 \quad 0
\end{array}\right]
$$%

Figure 1 gives the encoding circuit corresponding to the above operations. The above operations in reverse order take the unencoded stabilizer to the encoded stabilizer. And it is a $[[4,1,1]]_{5}$ qudit entanglement-assisted stabilizer quantum code.

(ii).The second example is a group $S$ generated by a non-commuting set of operators over $\mathbb{F}_{7}$ and its checking matrix is
$$
(X|Z)=\left[\begin{array}{c|c}
2 \quad 1 \quad 0 \quad 4 \quad 3 & 6 \quad 1 \quad 5 \quad 1 \quad 2\\
1 \quad 2 \quad 1 \quad 2 \quad 2 & 3 \quad 2 \quad 1 \quad 4 \quad 1\\
0 \quad 2 \quad 4 \quad 1 \quad 0 & 2 \quad 1 \quad 4 \quad 5 \quad 2\\
4 \quad 2 \quad 1 \quad 0 \quad 5 & 0 \quad 1 \quad 0 \quad 3 \quad 2
\end{array}\right]
$$
Use the encoding algorithm on the matrix, the matrix becomes
$$
(X|Z)=\left[\begin{array}{c|c}
1 \quad 0 \quad 0 \quad 0 \quad 0 & 0 \quad 0 \quad 0 \quad 0 \quad 0\\
0 \quad 0 \quad 0 \quad 0 \quad 0 & 1 \quad 0 \quad 0 \quad 0 \quad 0\\
0 \quad 1 \quad 0 \quad 0 \quad 0 & 0 \quad 0 \quad 0 \quad 0 \quad 0\\
0 \quad 0 \quad 0 \quad 0 \quad 0 & 0 \quad 2 \quad 0 \quad 0 \quad 0
\end{array}\right]
$$
So this non-commuting set of operators over $\mathbb{F}_{7}$ cannot construct an entanglement-assisted stabilizer quantum code.

\section{Conclusions}
In this paper, we present how to construct non-binary entanglement-assisted stabilizer quantum codes which do not require the dual-containing constraint and also give an algorithm to determine the circuit for non-binary entanglement-assisted stabilizer quantum codes. Many non-binary classical codes, like non-binary LDPC codes, which do not satisfy the dual-containing condition, can be used to construct non-binary entanglement-assisted stabilizer quantum codes. And the better the classical non-binary code is , the better the corresponding non-binary entanglement-assisted stabilizer quantum code will be.

\section*{Acknowledgments}
The authors acknowledge the support from the NNSF of China (Grant
No. 60403004) and the Outstanding Youth Foundation of Henan Province (Grant No. 0612000500).


\begin{thebibliography}{00}
\bibitem{01} P.~W.~Shor, Phys. Rev. A
\textbf{52}, R2493 (1995).

\bibitem{02}A.~M.~Steane, Proc. R. Soc. Lond. A.
\textbf{452}, 2521 (1996).

\bibitem{03} C.~H.~Bennett, D.~P.~DiVincenzo, J.~A.~Smolin and W.~K.~Wootters, Phys. Rev. A
\textbf{54}, 3824 (1996).

\bibitem{04} M.~Ohata and K.~Matsuura, e-print arXiv: 0702184.


\bibitem{05} D.~Gottesman, e-print arXiv: 0904.2557.


\bibitem{06} S.~Glancy, E.~Knill and H.~M.~Vasconcelos, Phys. Rev. A
\textbf{74}, 032319 (2006).

\bibitem{07} A.~R.~Calderbank, E.~M.~Rains, P.~W.~Shor and N.~J.~A.~Sloane, Phys. Rev. Lett.
\textbf{78}, 405 (1997).

\bibitem{08} A.~R.~Calderbank, E.~M.~Rains, P.~W.~Shor and N.~J.~A.~Sloane, IEEE Trans. Inf. Theory,
\textbf{44}, 1369 (1998).

\bibitem{09} D.~Gottersman, Phys. Rev. A
\textbf{54}, 1862 (1996).

\bibitem{10}  D.~Gottersman, Ph.D. dissertation, California Institute of Technology, Pasadena, CA,
 (1997).

\bibitem{11} M.~A.~Nielsen and I.~L.~Chuang,Cambridge, UK: Cambirdge University Press,
 (2000).

\bibitem{12} A.~R.~Calderbank and P.~W.~Shor, Phys. Rev. A
\textbf{54}, 1098 (1996).

\bibitem{13} G.~Bowen, Phys. Rev. A
\textbf{66}, 052313 (2002).

\bibitem{14} T.~Brun, I.~Devetak and M.~H.~Hsieh, Science
\textbf{314}, 436 (2006).

\bibitem{15} T.~Brun, I.~Devetak and M.~H.~Hsieh, e-print arXiv: 0608027.


\bibitem{16} M.~M.~Wilde, Ph.D. dissertation, University of Southern California,
 (2008).

\bibitem{17} E.~M.~Rains, IEEE Trans. Inf. Theory,
\textbf{106}, 080405 (1999).

\bibitem{18}D.~Gottesman, Chaos, Solitons, Fractals,
\textbf{10}, 1749 (1999)

\bibitem{19} A.~Ashikhmin and E.~Knill, IEEE Trans. Inf. Theory,
\textbf{47}, 3065 (2001).

\bibitem{20} M.~Grassl, M.~Rotteler, and T.~Beth, Internat. J. Found. Comput. Sci.,
\textbf{14}, 757 (2003).

\bibitem{21} A.~Ketkar, A.~Klappenecker, S.~Kumar, and P.~K.~Sarvepalli, IEEE Trans. Inf. Theory,
\textbf{51}, 4892 (2006).



\end{thebibliography}
\end{document}